\documentclass[sn-mathphys]{sn-jnl}

\usepackage{listings}
\usepackage{pmboxdraw}

\jyear{2023}%

\theoremstyle{thmstyleone}%
\theoremstyle{thmstyletwo}%

\theoremstyle{thmstylethree}%

\begin{document}

\title[ChatGPT for Semantic Database Management]{Context-based Ontology Modelling for Database: Enabling ChatGPT for Semantic Database Management}


\author[1,2]{\fnm{Wenjun} \sur{Lin}}\email{randy.lin@usask.ca}

\author[3]{\fnm{Paul} \sur{Babyn}}\email{paul.babyn@gmail.com}

\author[2]{\fnm{Yan} \sur{Yan}}\email{yyan@tru.ca}

\author*[1]{\fnm{Wenjun} \sur{Zhang}}\email{chris.zhang@usask.ca}

\affil*[1]{\orgdiv{College of Engineering}, \orgname{University of Saskatchewan}, \orgaddress{\street{57 Camous Dr}, \city{Saskatoon}, \postcode{S7N 5A9}, \state{SK}, \country{Canada}}}

\affil[2]{\orgdiv{College of Science}, \orgname{Thompson Rivers University}, \orgaddress{\street{835 University Dr}, \city{Kamloops}, \postcode{V2C 0C8}, \state{BC}, \country{Canada}}}



\abstract{This research paper explores the use of ChatGPT in database management. ChatGPT, an AI-powered chatbot, has limitations in performing tasks related to database management due to the lack of standardized vocabulary and grammar for representing database semantics. To address this limitation, the paper proposes a solution that involves developing a set of syntaxes that can represent database semantics in natural language. The syntax is used to convert database schemas into natural language formats, providing a new application of ChatGPT in database management.

The proposed solution is demonstrated through a case study where ChatGPT is used to perform two tasks, semantic integration, and tables joining. Results demonstrate that the use of semantic database representations produces more precise outcomes and avoids common mistakes compared to cases with no semantic representation.

The proposed method has the potential to speed up the database management process, reduce the level of understanding required for database domain knowledge, and enable automatic database operations without accessing the actual data, thus illuminating privacy protection concerns when using AI. This paper provides a promising new direction for research in the field of AI-based database management.
}

\keywords{ChatGPT, Database management, Semantic database representation, Semantic integration, Tables joining}



\maketitle

\section{Introduction}\label{sec1}

ChatGPT is a conversational chatbot that uses artificial intelligence (AI) and machine learning (ML) techniques, combined with natural language processing (NLP) methods, to produce human-like text. It was launched in November 2022 and quickly gained popularity, reaching over one million users within just five days \cite{thorp2023chatgpt}. ChatGPT's ability to produce human-like text and perform a wide range of tasks has made it a popular tool for many users, including answering questions, writing short stories, composing music, solving math problems, performing language translations, and even computer programming.

Database operation involves the manipulation of data and information using specific syntax or commands, similar to computer programming. In database operations, these commands are known as database queries, which are used to retrieve, update, and manipulate data stored in a database. Similarly, in computer programming, instructions are written in a programming language, such as C or Python, in order to specify the desired behaviour of a computer program.

There has been an expectation that ChatGPT could assist in creating database queries, just as it can assist in creating computer programs. However, creating database queries requires an understanding of the database itself, and there is no conventional way to represent database semantics. This problem limits ChatGPT's ability to perform tasks related to database management.

In this paper, we present a solution to this problem by developing a set of syntax that can represent database semantics, such as table structure and relationships, in natural language. This allows for the creation of semantic representations of databases that can be understood by ChatGPT and enable it to perform database management tasks. Our work is demonstrated through a case study, where ChatGPT is used to perform two tasks: semantic integration and table joining. Our results show that the use of semantic database representations produces more precise outcomes and avoids common mistakes compared to cases with no semantic representation.

The proposed method transforms database schemas into natural language formats, providing a new application of ChatGPT in database management. This study has the potential to speed up the database management process, reduce the level of understanding required for database domain knowledge, and enable automatic database operations without accessing the actual data, thus illuminating privacy protection concerns when using AI.

The rest of the paper is organized as follows: In Section \ref{sec2}, we provide a review of related work in the area of database management using AI. Then, we describe our proposed solution in Section \ref{sec3}. Section \ref{sec4} presents the results and discussion of our case study. Finally, we discuss the potential benefits and limitations of our work and conclude with future directions for research.

\section{Literature review}\label{sec2}
\subsection{AI-based database queries generation}

The use of AI models for generating database queries through natural language has been the focus of several research studies. One such model proposed by Bais et al. \cite{bais2016model} utilizes NLP techniques to analyze and interpret user queries by performing morphological, syntactic, and semantic analysis, resulting in a valid database query in SQL. Similarly, Sawant et al. \cite{sawant2022ai} implemented a system that can generate SQL queries from text and speech input using NLP and deep learning techniques such as Long Short Term Memory (LSTM).

Other studies, such as Ghosh et al. \cite{ghosh2014automatic}, Nagare et al. \cite{nagare2017automatic}, and Kombade et al. \cite{kombade2020natural} , have also utilized techniques such as lexical analysis, syntax analysis, and semantic analysis to extract SQL queries from natural language input. Kombade et al. \cite{kombade2020natural} even considered the use of abbreviations in NLP to generate SQL queries. The implementation of these studies used python with a GUI for input and output, and the user could provide input through speech or text.

Despite the progress made in this field, limitations still exist in the ability of AI models to accurately generate database queries from natural language due to the complexity and ambiguity of natural language, as well as the lack of standardized vocabulary and grammar for representing database structures. For instance, Nagare et al. \cite{nagare2017automatic} mentions that the system checks the validity of the user's query, but it is unclear how the query's validity is determined. Moreover, the studies only consider basic database operations such as select, delete, and update. Complex operations, such as joining multiple tables and semantic integration, have not been investigated.

\subsection{Semantic integration}

Semantic integration is crucial for resolving mismatches in data representation between related databases. In today's digital world, organizations are generating and storing vast amounts of data in various databases, which often leads to inconsistencies in the data representation. For instance, one database may use the attribute "Social Security Number" to identify individuals, while another database may use the attribute "SSN". In such cases, it is crucial to determine the relationships between these attributes in order to accurately compare and combine the data from these databases. 

Several approaches for semantic integration have been developed in recent years. Tools like Silk \cite{volz2009silk}, LIMES \cite{ngomo2011limes}, and PARIS \cite{suchanek2011paris} use string similarity metrics, functional properties, and manual configuration to detect matching attributes. WebPie \cite{urbani2012webpie} and LINDA \cite{bohm2012linda} are fully automatic systems that use techniques such as neighborhood checks and block placement. MateTee \cite{morales2017matetee} and RDF2VEC \cite{ristoski2016rdf2vec} are more recent approaches that utilize embeddings and machine learning to find similarities. However, the task of matching entities can become complicated with the presence of entities in abbreviates \cite{lantzaki2017radius}. 

The existing approaches are heavily relying on domain expertise and require complex preparations. For example, dataset-based approaches \cite{morales2017matetee, ristoski2016rdf2vec} require determining the relationships among attributes. Keyword-based approaches \cite{volz2009silk,ngomo2011limes} depend on the accuracy of metadata, while URI-based approaches \cite{urbani2012webpie} strongly depend on dereferencing HTTP URIs. 

\section{Methodology}\label{sec3}

\subsection{ChatGPT }

ChatGPT is a language model developed by OpenAI \cite{van2023chatgpt}. It is a type of AI algorithm trained to predict the likelihood of a given sequence of words based on the context of the words that come before it. This technology is based on self-attention mechanisms \cite{humphreys2016attentional} and has been trained on a massive dataset of text, allowing it to generate sophisticated and seemingly intelligent writing. ChatGPT is designed to converse with users in English and other languages on a wide range of topics, making it ideal for use in chatbots, customer service, content creation, and language translation tasks. 

One of the applications of ChatGPT is to assist in programming, which can be achieved in two ways. Firstly, ChatGPT can serve as a programming assistant or tool. For instance, developers can ask ChatGPT programming-related questions and obtain recommendations and suggestions about general workflows and steps. Secondly, ChatGPT can generate code snippets directly, resulting in enhanced productivity and time-saving benefits for developers.

Despite its advanced natural language processing capability and successes in assisting programming, ChatGPT has not yet been able to generate queries for databases because database schemas, which contain vital information about database structures, are frequently written in the form of a graph rather than natural language.

\subsection{Context-based Ontology Modelling for Database}

This study presents a new method, Context-based Ontology Modelling for Database (COM-DB), which is aimed at converting database schemas into natural language. COM-DB is built upon our previous research on ontology modelling \cite{lin2023ontology}, which utilizes constructs like context-of, monodirectional relationship, and bi-directional relationship to describe the relationship of concepts in databases. 

In this study, we focus on the usage of the “context-of” construct for describing database schema, especially at the conceptual data model level. A conceptual schema or conceptual data model is a map of concepts and their relationships used for databases. The key feature of COM-DB is the ability to convert these relationships into natural language, which makes it more accessible to ChatGPT. To demonstrate the effectiveness of COM-DB, we provide two examples that illustrate how the “context-of” construct can be used to describe the relationship of headers within one table and the relationship of tables within one database. 

\begin{figure}[h]%
\centering
\includegraphics[width=0.5\textwidth]{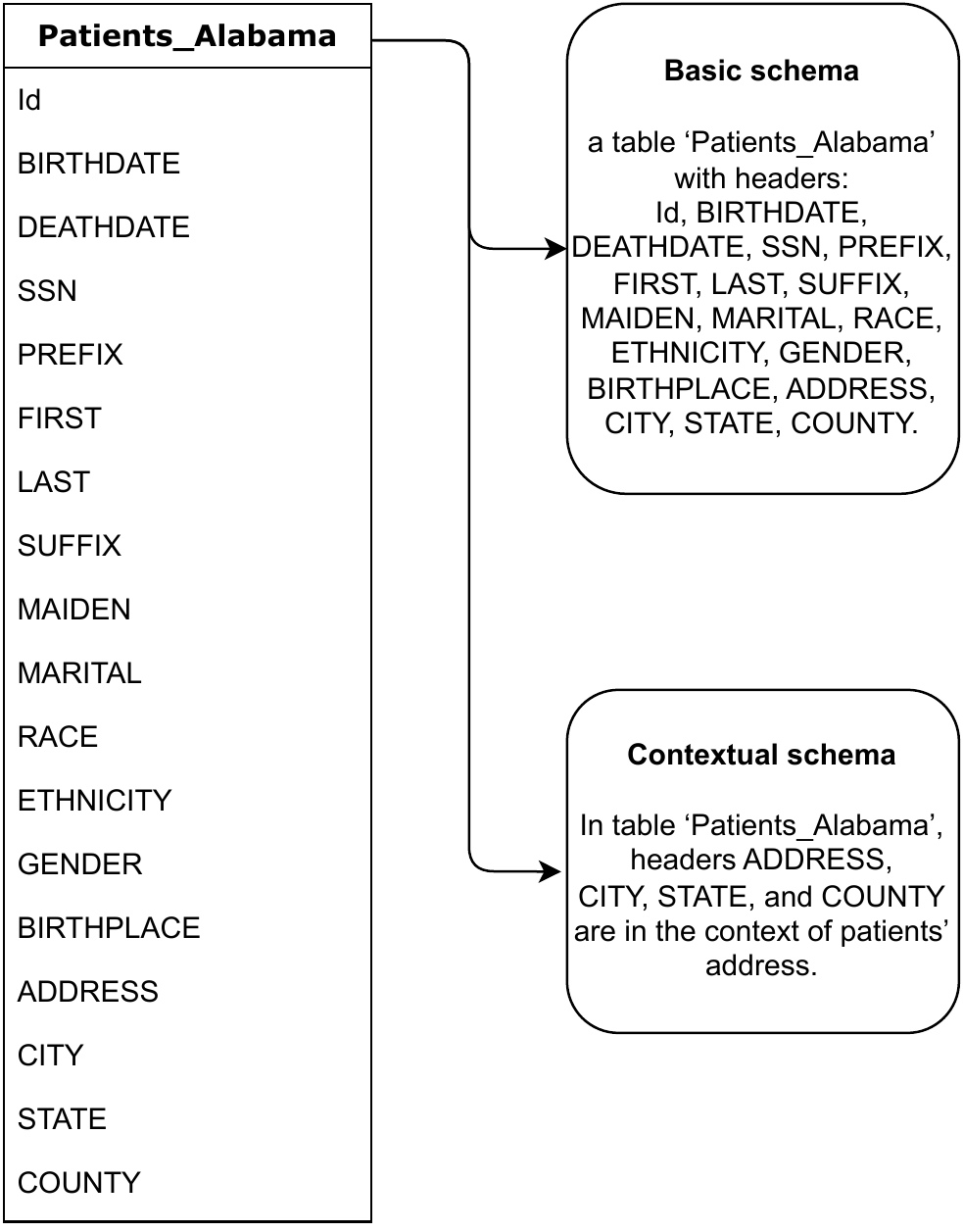}
\caption{Describing relationships of headers within one table using COM-DB}\label{fig1}
\end{figure}

The example in Figure \ref{fig1} demonstrates how COM-DB can be used to describe the relationships between headers within a database table. The left side of the figure shows a typical database table schema, which includes the name of the table and the names of its headers.

COM-DB converts this schema into two parts: the base schema, which describes the names of the headers in the table, and the contextual schema, which uses the “context-of” construct to describe relationships among headers. In this example, the headers "ADDRESS", "CITY", "STATE", and "COUNTY" are related and are used to store information about patients' addresses.

The contextual schema condenses this information into a single sentence by using the “context-of” construct to describe the relationship between these headers and the patients' addresses. Specifically, it states that "ADDRESS, CITY, STATE, COUNTY are in the context of patients' address". This represents four relationships in one sentence, making the information more easily understandable and less verbose.

\begin{figure}[h]%
\centering
\includegraphics[width=0.9\textwidth]{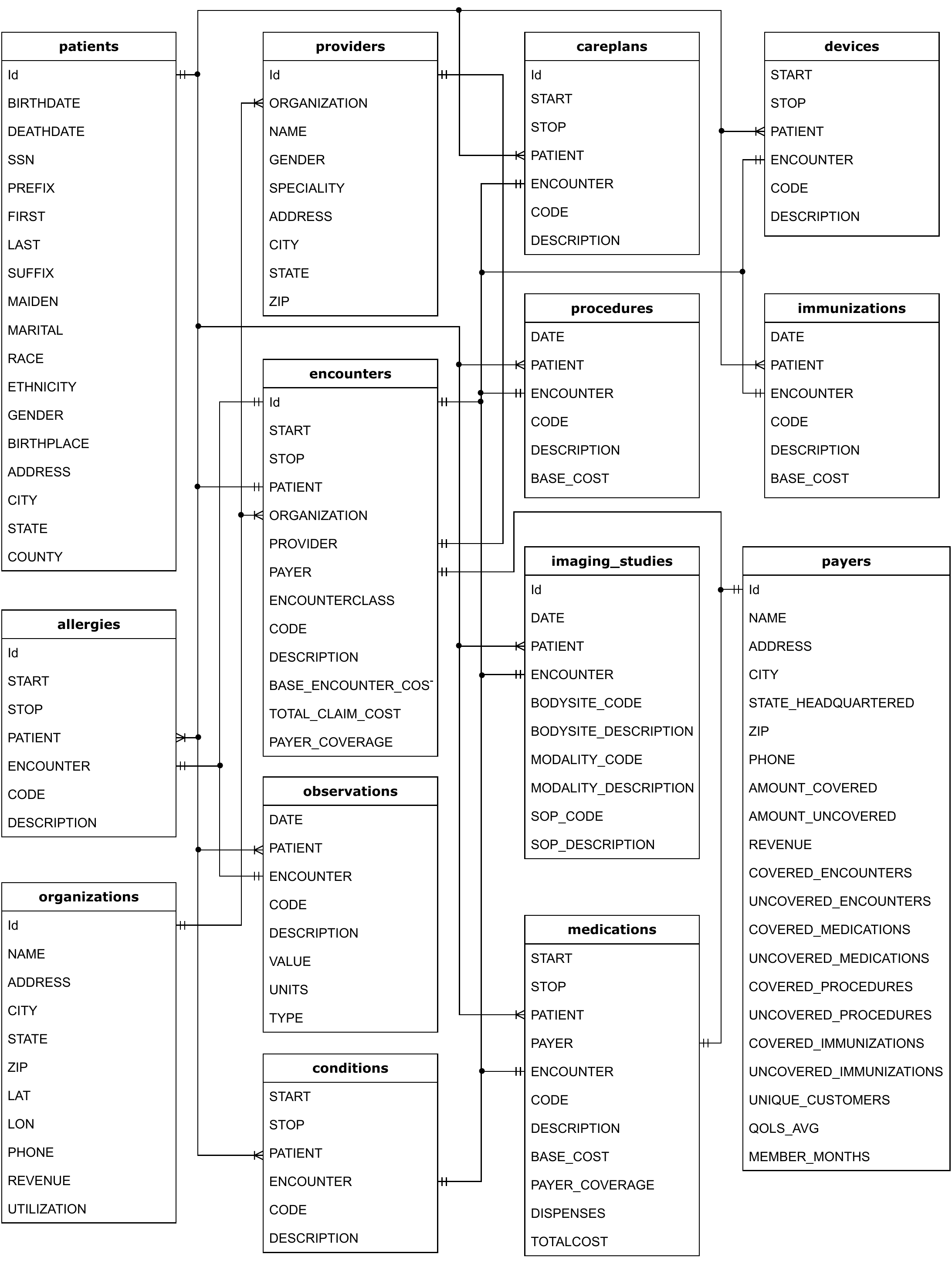}
\caption{Entity-Relationship schema of a hospital database \cite{walonoski2018synthea}}\label{fig2}
\end{figure}

Figure \ref{fig2} is an Entity-Relationship schema of a hospital database. Similar to the first example, COM-DB represents the database in two parts. The first part, basic schema, explain headers in each table. Some basic schemas are as follows:

\begin{lstlisting}
Given a table 'allergies' with headers: Id, START, STOP, PATIENT, ENCOUNTER, CODE, DESCRIPTION. 
And a table 'careplans' with headers: Id, START, STOP, PATIENT, ENCOUNTER, CODE, DESCRIPTION. 
And a table 'conditions' with headers: START, STOP, PATIENT, ENCOUNTER, CODE, DESCRIPTION. 
And a table 'devices' with headers: START, STOP, PATIENT, ENCOUNTER, CODE, DESCRIPTION, UDI. 
And a table 'encounters' with headers: Id, START, STOP, PATIENT, ORGANIZATION, PROVIDER, PAYER, ENCOUNTERCLASS, CODE, DESCRIPTION, BASE_ENCOUNTER_COST, TOTAL_CLAIM_COST, PAYER_COVERAGE. 
And a table 'imaging_studies' with headers: Id, DATE, PATIENT, ENCOUNTER, BODYSITE_CODE, BODYSITE_DESCRIPTION, MODALITY_CODE, MODALITY_DESCRIPTION, SOP_CODE, SOP_DESCRIPTION. 
And a table 'immunizations' with headers: DATE, PATIENT, ENCOUNTER, CODE, DESCRIPTION, BASE_COST. 
And a table 'medications' with headers: START, STOP, PATIENT, PAYER, ENCOUNTER, CODE, DESCRIPTION, BASE_COST, PAYER_COVERAGE, DISPENSES, TOTALCOST. 
And a table 'observations' with headers: DATE, PATIENT, ENCOUNTER, CODE, DESCRIPTION, VALUE, UNITS, TYPE. 
And a table 'organizations' with headers: Id, NAME, ADDRESS, CITY, STATE, ZIP, LAT, LON, PHONE, REVENUE, UTILIZATION. 
And a table 'patients' with headers: Id, BIRTHDATE, DEATHDATE, SSN, PREFIX, FIRST, LAST, SUFFIX, MAIDEN, MARITAL, RACE, ETHNICITY, GENDER, BIRTHPLACE, ADDRESS, CITY, STATE, COUNTY. 
And a table 'payers' with headers: Id, NAME, ADDRESS, CITY, STATE_HEADQUARTERED, ZIP, PHONE, AMOUNT_COVERED, AMOUNT_UNCOVERED, REVENUE, COVERED_ENCOUNTERS, UNCOVERED_ENCOUNTERS, COVERED_MEDICATIONS, UNCOVERED_MEDICATIONS, COVERED_PROCEDURES, UNCOVERED_PROCEDURES, COVERED_IMMUNIZATIONS, UNCOVERED_IMMUNIZATIONS, UNIQUE_CUSTOMERS, QOLS_AVG, MEMBER_MONTHS. 
And a table 'procedures' with headers: DATE, PATIENT, ENCOUNTER, CODE, DESCRIPTION, BASE_COST. 
And a table 'providers' with headers: Id, ORGANIZATION, NAME, GENDER, SPECIALITY, ADDRESS, CITY, STATE, ZIP.
\end{lstlisting}

In addition, the second part, contextual schema, is as follows:

\begin{lstlisting}
allergies, careplans, conditions, devices, immunizations, observations, procedures, imaging_studies are in the context of patients, encounters. 
encounters are in the context of patients, organizations, providers, payers. 
medications are in the context of patients, encounters, payers. 
providers are in the context of organizations.
\end{lstlisting}

Note that the contextual schema is shown in a condensed form. For example, “allergies, careplans, conditions, devices, immunizations, observations, procedures, imaging\_studies are in the context of patients, encounters.” represents 8 × 2 = 16 relationships. These relationships are between 8 tables “allergies, careplans, conditions, devices, immunizations, observations, procedures, imaging\_studies” and 2 tables “patients, encounters”. 

We use these examples to show practical applications of COM-DB and how it can be used to generate natural language descriptions of complex database schemas. Overall, the methodology of this study involves designing and implementing the COM-DB method, which includes utilizing the “context-of” construct and other ontology modelling constructs to convert database schema into natural language. The effectiveness of this method is demonstrated through the use of two examples, which show how it can be used to complete two sophisticated database management tasks. The effectiveness of the method is demonstrated in the case study. 

\section{Case study}\label{sec4}

The case study aims to showcase the efficacy of the proposed COM-DB system. The system's primary feature is the “context-of” construct, which utilizes natural language to capture database semantics like table structure and relationships. The primary objective of the system is to create semantic representations of databases that can be easily comprehended by ChatGPT, enabling it to perform various database management tasks.

The case study provides empirical evidence to support the effectiveness of COM-DB. Two sample databases are collected from the literature, Synthea\_Alabama \cite{walonoski2018synthea} and BDA\_EHR \cite{silvestri2019big}. Based on those databases, two experiments are conducted that represent typical tasks conducted during database integration: semantic integration and tables joining. In both experiments, ChatGPT is used to perform tasks with and without the COM-DB-based schema. The study repeats each experiment 10 times to ensure reliability and eliminate the potential inconsistency in ChatGPT's performance. Results demonstrated illustrate an average result from the repeated experiments. 

\subsection{Experiment 1: Semantic Integration}

Semantic integration involves merging two tables with the same category of information. Different headers' names among different tables often cause incompatibility issues. 'patients\_A' and 'patients\_B' are tables of patient information from BDA\_EHR and Synthea\_Alabama, respectively. 'patients\_A' contains headers: Id\_patients, Name, Surname, Date of Birth, Place of Birth, Address, Gender, Blood Type, Job. And 'patients\_B' contains headers: Id, BIRTHDATE, DEATHDATE, SSN, PREFIX, FIRST, LAST, SUFFIX, MAIDEN, MARITAL, RACE, ETHNICITY, GENDER, BIRTHPLACE, ADDRESS, CITY, STATE, COUNTY.

The goal of this experiment is to identify headers from table 'patients\_A' and table 'patients\_B' which contain the same information, knowing that some headers may need to be combined or split for the mapping. The ideal mapping is illustrated in Table \ref{tab1}.

\begin{table}[h]
\begin{center}
\begin{minipage}{174pt}
\caption{The ideal header mapping results of table 'patients\_A' and table 'patients\_B'}\label{tab1}%
\begin{tabular}{@{}ll@{}}
\toprule
patients\_A & patients\_B\\
\midrule
Name                 & FIRST                                                                                                                                  \\
Surname              & LAST                                                                                                                                   \\
Date of Birth        & BIRTHDATE                                                                                                                              \\
Place of Birth       & BIRTHPLACE                                                                                                                             \\
Address              & ADDRESS CITY STATE COUNTY                             \\
Gender               & GENDER                                                \\    \botrule
\end{tabular}
\end{minipage}
\end{center}
\end{table}

\begin{figure}[h]%
\centering
\includegraphics[width=0.5\textwidth]{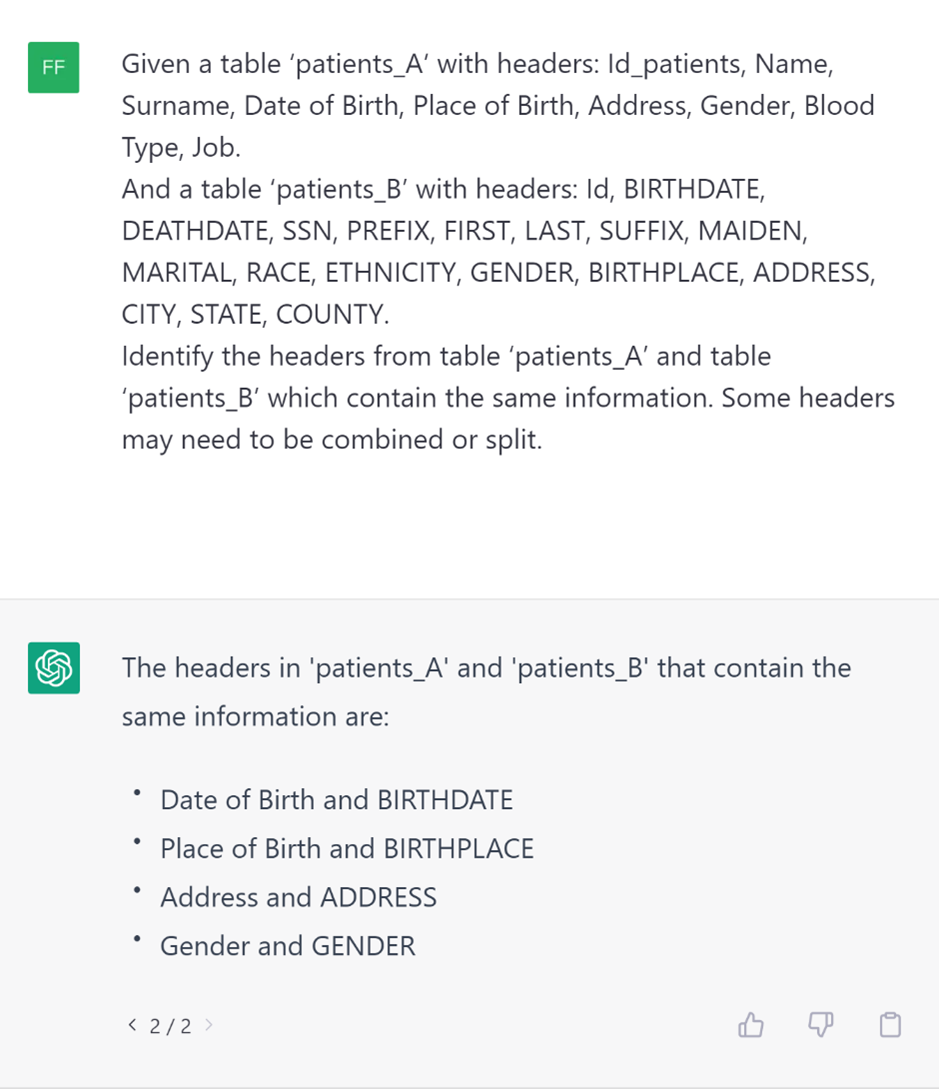}
\caption{Experiment 1, header mapping without COM-DB based schema}\label{fig3}
\end{figure}

Figure \ref{fig3} shows the input and output of using ChatGPT without COM-DB based schema. The message from the icon 'FF' is the input, while the message from the graphical icon is the output from ChatGPT. The input contains two parts of information. The first part is to explain the situation, which contains the names of headers in each table. Note that only the headers are provided here, without any sample data or data type. The second part “Identify the headers from table 'patients\_A' and table 'patients\_B' which contain the same information. Some headers may need to be combined or split.” is an explanation of the task to be completed by ChatGPT.

The output in Figure \ref{fig3} shows that ChatGPT can understand the task and perform it to a degree. It matches Date of Birth and BIRTHDATE, Place of Birth and BIRTHPLACE, Gender and GENDER, correctly. However, it failed to match Name with FIRST, and Surname with LAST. In addition, ADDRESS in 'patients\_B' should be used with other headers CITY, STATE, COUNTY. This was not noticed by ChatGPT.

\begin{figure}[h]%
\centering
\includegraphics[width=0.5\textwidth]{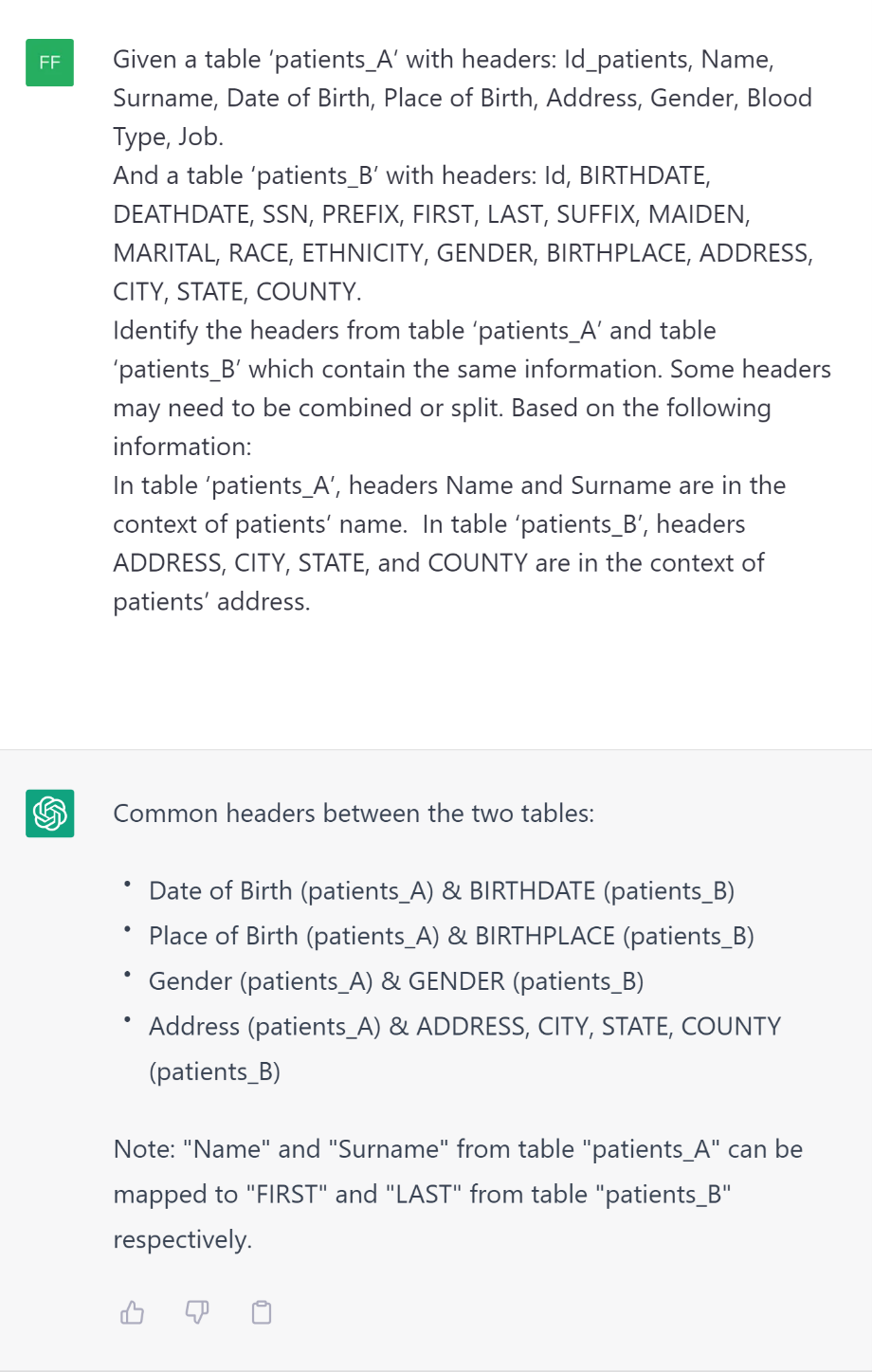}
\caption{Experiment 1, header mapping with COM-DB based schema}\label{fig4}
\end{figure}

Figure \ref{fig4} shows the input and output of using ChatGPT with COM-DB based schema. In addition to the inputs used in Figure \ref{fig3}, the ontology model information is described as “In table 'patients\_A', headers Name and Surname are in the context of patients' name. In table 'patients\_B', headers ADDRESS, CITY, STATE, and COUNTY are in the context of patients' address.”

Figure \ref{fig4} illustrates a typical result from ChaGPT and demonstrates the COM-DB based schema improves the performance of semantic integration as ChatGPT has successfully identified all mappings as expected.

\subsection{Experiment 2: Tables Joining}

Tables joining involves generating a new table or view that combines data from multiple tables. This process requires an Entity Relationship schema which doesn't have a conventional way to be represented in natural language. To demonstrate the effectiveness of COM-DB based schema in representing the Entity Relationship, Synthea\_Alabama is used in this experiment. The database contains 14 tables as shown in Figure \ref{fig2}. The goal of the experiment is to create a SQL query that generates a list of careplans, with corresponding providers' and patients' identity information. The careplans are advices from providers (such as physicians) to patients. A SQL query needs to properly join four tables: careplans, providers, patients, and encounters. The encounters table plays a critical role here as it connects the patients table with the careplans table. This information is typically contained in an Entity Relationship schema. 

\begin{figure}[h]%
\centering
\begin{minipage}{.5\textwidth}
  \centering
  \includegraphics[width=0.9\textwidth]{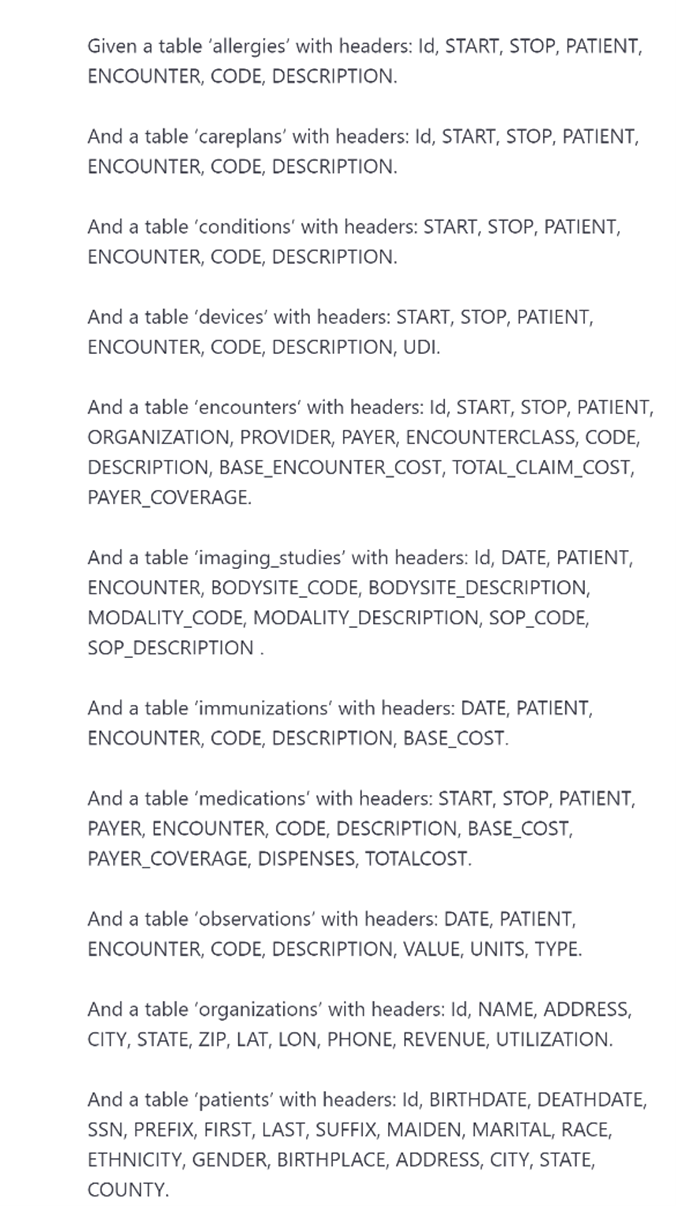}
\end{minipage}%
\begin{minipage}{.5\textwidth}
  \centering
  \includegraphics[width=0.9\textwidth]{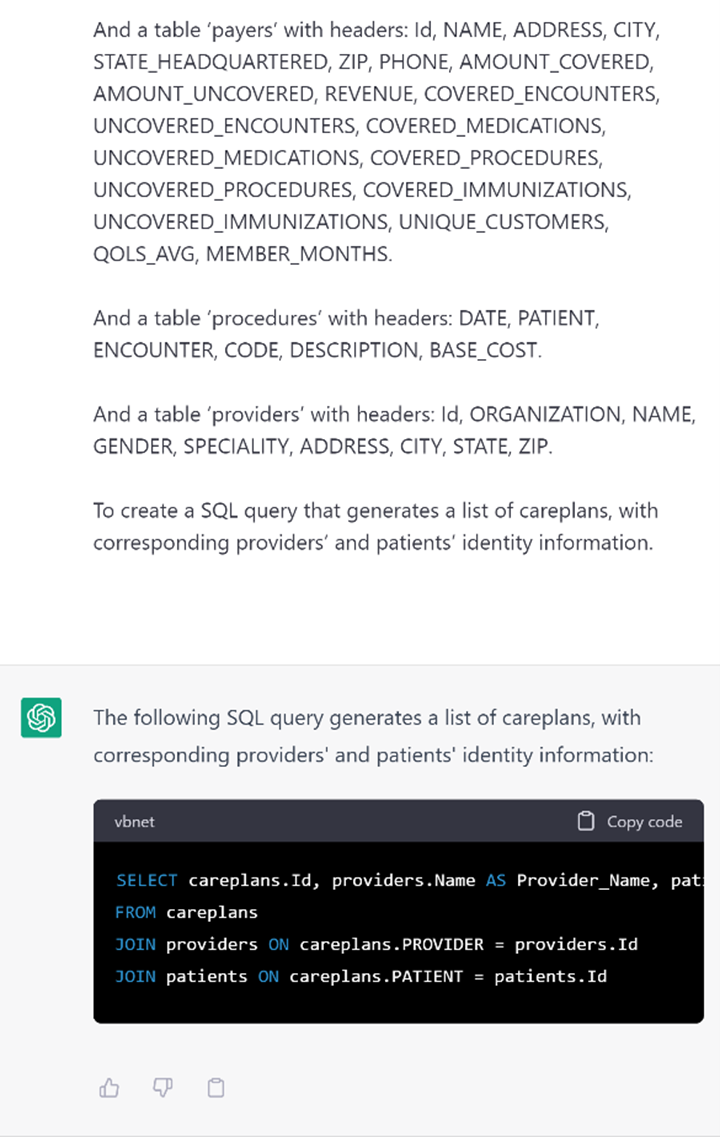}
\end{minipage}
\caption{Experiment 2 Generate a new view from multiple tables without COM-DB based schema. The conversation is split into two columns, from left to right.}\label{fig5}
\end{figure}

Figure \ref{fig5} shows the input and output of using ChatGPT without COM-DB based schema. The input contains two parts of information. The first part is to explain all tables with their contained headers in alphabetical order. Similar to Experiment 1, only the headers are provided here without any sample data or data type. The second part, “To create a SQL query that generates a list of careplans, with corresponding providers' and patients' identity information.” is an explanation of the task to be completed by ChatGPT.

\begin{figure}[h]%
\centering
\includegraphics[width=0.73\textwidth]{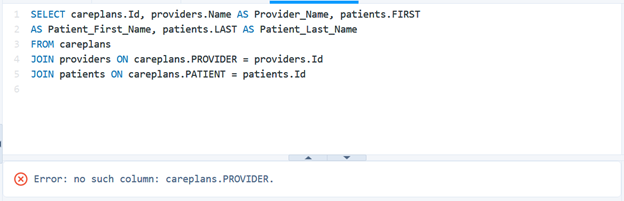}
\caption{Experiment 2, SQL query results without COM-DB based schema}\label{fig6}
\end{figure}

The output from ChatGPT is verified by executing the SQL query in the hospital database. The result is shown in Figure \ref{fig6}. From the result, it was found the query doesn't work due to the error “no such column: careplans.PROVIDER”. The root cause of this error is the missing the encounters table as explained earlier.

\begin{figure}[h]%
\centering
\begin{minipage}{.5\textwidth}
  \centering
  \includegraphics[width=0.9\textwidth]{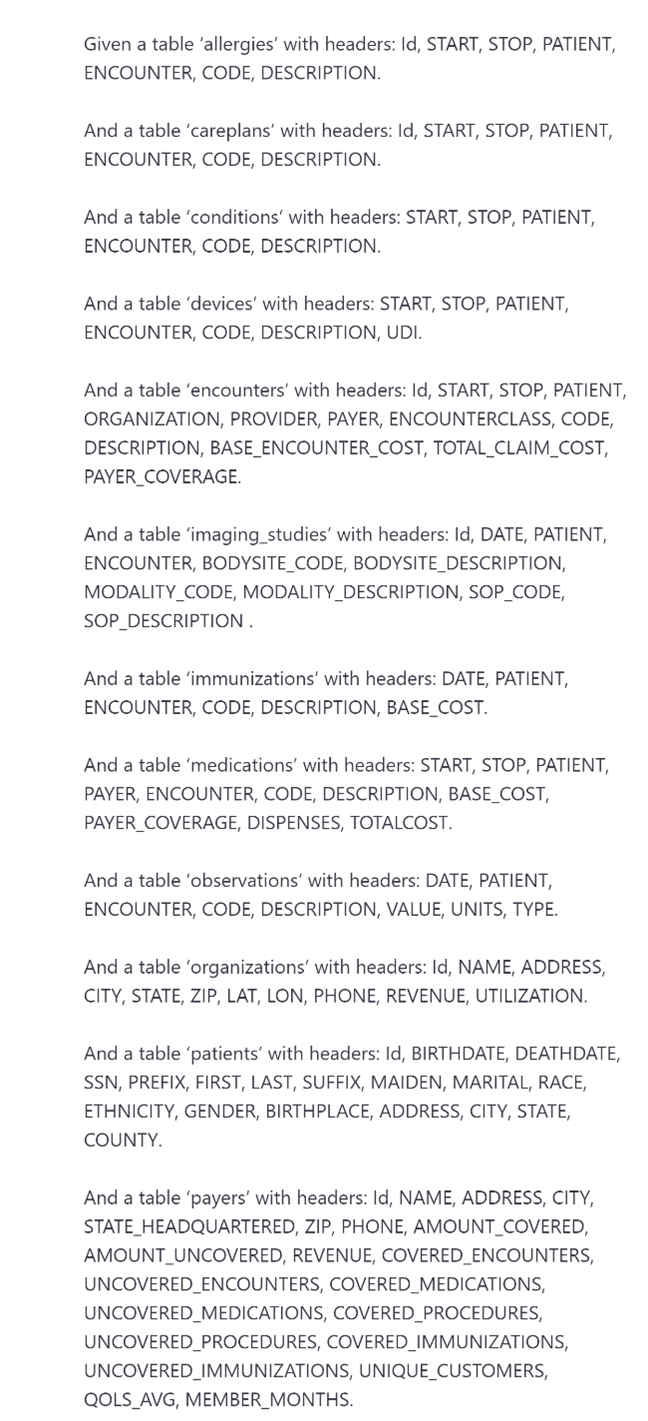}
\end{minipage}%
\begin{minipage}{.5\textwidth}
  \centering
  \includegraphics[width=0.9\textwidth]{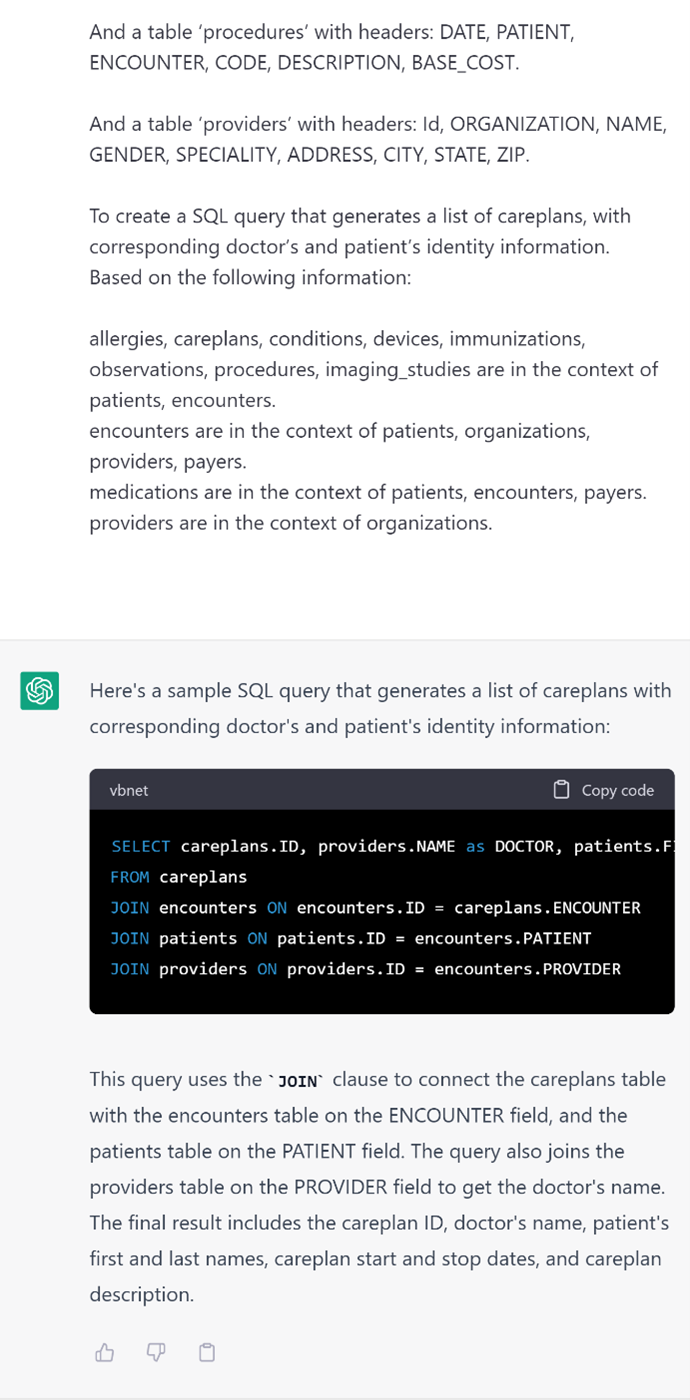}
\end{minipage}
\caption{Experiment 2 Generate a new view from multiple tables with COM-DB based schema. The conversation is split into two columns, from left to right.}\label{fig7}
\end{figure}

\begin{figure}[h]%
\centering
\includegraphics[width=0.73\textwidth]{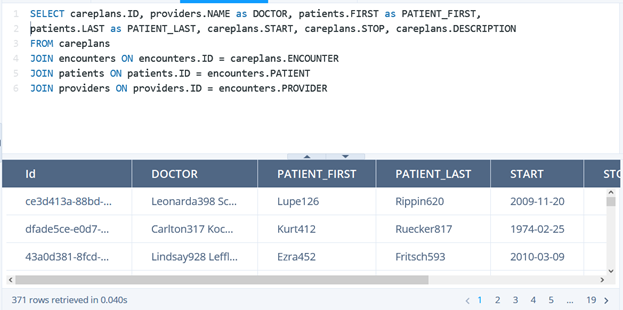}
\caption{Experiment 2, SQL query results with COM-DB based schema}\label{fig8}
\end{figure}

Figure \ref{fig7} shows the input and output of using ChatGPT with COM-DB based schema that explains the context information of each table. In this case, the COM-DB based schema describes relations between tables. Figure \ref{fig8} verifies the SQL query by executing it in the hospital database. It shows that ChatGPT has successfully generated the query that results in a correct view. 

\section{Discussions}\label{sec5}

The results of the experiments indicate that ChatGPT performs better in both semantic integration and tables joining tasks when using the COM-DB-based schema. The context information provided by the ontology models helps ChatGPT to better complete the tasks. The study demonstrates that the “context-of” construct in COM-DB captures certain context information which may not be included in a conventional ontology model. This context information determines the relations between concepts, which helps to eliminate ambiguities between concepts and increases the chances of success during database integration.

In addition, COM-DB enables automatic database operations without compromising privacy protections. Unlike existing AI-enabled database operations, which require the AI algorithm to access all data in the database, COM-DB only generates a schema from the table structure, instead of its content. This schema does not contain any privacy information, making it safe to share with third-party services such as ChatGPT. By sending the schema to ChatGPT, the risk of privacy breach is significantly reduced, as ChatGPT can perform automated database operations without ever accessing sensitive data. This allows businesses and organizations to leverage the power of AI and automation to streamline their operations and improve efficiency, without sacrificing the privacy and security of their customers' data. With COM-DB, businesses can have peace of mind knowing that their data is secure and protected, while still enjoying the benefits of automated database operations.

\section{Conclusion}\label{sec6}
This paper explores the use of ChatGPT in the area of database management, highlighting the challenges of using natural language processing to perform database queries. Our research presents a solution by developing a set of syntaxes to represent database semantics in natural language. These syntaxes, called COM-DB, enable ChatGPT to perform tasks related to database management, such as semantic integration and tables joining. Our case study shows that the use of semantic representations in database management leads to more precise outcomes and reduces common mistakes compared to cases without such representations.

Our research aims to contribute to the field of database management by introducing a novel approach for converting database schemas into natural language format, thereby opening up new applications for ChatGPT. This approach has the potential to deliver significant benefits, including faster database management, reduced domain knowledge requirements, and enhanced privacy protection through automated database operations that do not require access to actual data.

Future work involves expanding the scope of our method to include more complex database operations and testing it on larger databases. Furthermore, we intend to investigate the feasibility of incorporating other natural language processing models into database management and explore the possibilities of combining various models to enhance their capabilities.

In conclusion, our research demonstrates the potential of natural language processing models to be employed in the field of database management, providing a new way to interact with and manipulate databases. By leveraging the power of ChatGPT alongside our COM-DB syntaxes, we have demonstrated that complex database operations can be executed using natural language, offering a new approach to simplify database management and enhance productivity.

\bmhead{Acknowledgments}

This study is supported by Natural Sciences and Engineering Research Council of Canada, Alliance grant \#ALLRP 555161 - 20.





\bibliography{sn-bibliography}


\end{document}